\documentclass[journal=jacsat,manuscript=article]{achemso}

\usepackage[version=3]{mhchem} 

\usepackage{xcolor}
\definecolor{DragonGreen}{RGB}{0,126,48}

\newcommand{\revision}[1]{{\color{black} #1 }}

\usepackage{listings}
\usepackage{xparse}
\usepackage{gensymb}

\NewDocumentCommand{\codeword}{v}{%
\texttt{\textcolor{black}{#1}}%
}

\newcommand{\abs}[1]{\lvert#1\rvert}%
\newcommand{\norm}[1]{\left\lVert#1\right\rVert}

\author{Piero Gasparotto}%
\altaffiliation{Authors contributed equally to the manuscript}
\email{piero.gasparotto@empa.ch}
\affiliation{%
Empa, Swiss Federal Laboratories for Materials Science and Technology, nanotech@surfaces Laboratory, Überlandstrasse 129, 8600 Dübendorf, Switzerland
}%
\author{Maria Fischer}
\altaffiliation{Authors contributed equally to the manuscript}
\affiliation{Empa, Swiss Federal Laboratories for Materials Science and Technology, Laboratory for Magnetic and Functional Thin Films, Überlandstrasse 129, 8600 Dübendorf, Switzerland}%
\author{Daniele Scopece}%
\affiliation{%
Empa, Swiss Federal Laboratories for Materials Science and Technology, nanotech@surfaces Laboratory, Überlandstrasse 129, 8600 Dübendorf, Switzerland
}%
\author{Maciej O. Liedke}
\affiliation{Helmholtz-Zentrum Dresden-Rossendorf, Institute of Radiation Physics, Bautzner Landstrasse 400, 01328 Dresden, Germany}
\author{Maik Butterling}
\affiliation{Helmholtz-Zentrum Dresden-Rossendorf, Institute of Radiation Physics, Bautzner Landstrasse 400, 01328 Dresden, Germany}
\author{Andreas Wagner}
\affiliation{Helmholtz-Zentrum Dresden-Rossendorf, Institute of Radiation Physics, Bautzner Landstrasse 400, 01328 Dresden, Germany}
\author{Oguz Yildirim}
\affiliation{Empa, Swiss Federal Laboratories for Materials Science and Technology, Laboratory for Magnetic and Functional Thin Films, Überlandstrasse 129, 8600 Dübendorf, Switzerland}
\author{Mathis Trant}
\affiliation{Empa, Swiss Federal Laboratories for Materials Science and Technology, Laboratory for Magnetic and Functional Thin Films, Überlandstrasse 129, 8600 Dübendorf, Switzerland}
\author{Daniele Passerone}
\affiliation{%
Empa, Swiss Federal Laboratories for Materials Science and Technology, nanotech@surfaces Laboratory, Überlandstrasse 129, 8600 Dübendorf, Switzerland
}%
\author{Hans J. Hug}
\email{hans-josef.hug@empa.ch}
\affiliation{Empa, Swiss Federal Laboratories for Materials Science and Technology, Laboratory for Magnetic and Functional Thin Films, Überlandstrasse 129, 8600 Dübendorf, Switzerland}
\alsoaffiliation{University of Basel, Department of Physics, Klingelbergstrasse 82, 4056 Basel, Switzerland}
\author{Carlo A. Pignedoli}
\email{carlo.pignedoli@empa.ch}
\affiliation{%
Empa, Swiss Federal Laboratories for Materials Science and Technology, nanotech@surfaces Laboratory, Überlandstrasse 129, 8600 Dübendorf, Switzerland
}%

\title[]{Mapping the Structure of Oxygen-Doped Wurtzite Aluminum Nitride Coatings From Ab Initio Random Structure Search and Experiments}

\abbreviations{PALS,XRD,ML,KPCA,PCA,DFT}

\begin{document}








\revision{\noindent\emph{\textbf{Keywords}: Unsupervised learning, ab initio random structure search, SOAP descriptor, kernel PCA, oxygen-doped wurtzite alloys, lattice defects, X-ray diffraction, PALS, DFT.}}
\newpage

\begin{abstract}
Machine learning is changing how we design and interpret experiments in materials science. In this work we show how unsupervised learning, combined with \emph{ab initio} random structure searching, improves our understanding of structural metastability in multicomponent alloys. We focus on the case of Al-O-N alloys where the formation of aluminum vacancies in wurtzite AlN upon the incorporation of substitutional oxygen can be seen as a general mechanism of solids where crystal symmetry is reduced to stabilize defects. The ideal AlN wurtzite crystal structure occupation cannot be matched due to the presence of an aliovalent hetero-element into the structure. The traditional interpretation of the c-lattice shrinkage in sputter-deposited Al-O-N films from X-ray diffraction (XRD) experiments suggests the existence of a solubility limit at 8$at.\%$ oxygen content. Here we show that such naive interpretation is misleading. 
We support XRD data with accurate \emph{ab initio} modeling and dimensionality reduction on advanced structural descriptors to map structure-property relationships. No signs of a possible solubility limit are found. Instead, the presence of a wide range of non-equilibrium oxygen-rich defective structures emerging at increasing oxygen contents suggests that the formation of grain boundaries is the most plausible mechanism responsible for the lattice shrinkage measured in Al-O-N sputtered films.  We further confirm our hypothesis using positron annihilation lifetime spectroscopy.
\end{abstract}

\section{Introduction}
Nanocomposite hard coatings are an important class of materials for various applications. Ti-Si-N, Al-Si-O\cite{9,10}, and Al-O-N\cite{13} thin films fabricated by reactive sputtering are typical examples. Because of the wide band-gap of AlN, the last two materials are optically transparent in addition to their high hardness. Their properties, such as the strain of the films, the index of refraction and their hardness can be tuned over a wide range by changing their chemical composition, i.e. by the amount of Si and O added. 

For Al-O-N, Harris et al. have shown that up to $0.75\%$ O can be incorporated into the wurtzite lattice of single crystalline materials synthesized under thermal equilibrium conditions\cite{5}. Oxygen has a high Pauling electronegativity of $\chi_{\mathrm{O}} = 3.5$ as does the electron ($e^-$) acceptor N with $\chi_{\mathrm{N}} = 3.0$\cite{6}. Slack and coworkers therefore proposed that O substitutes the N (O$_{\rm N}$) in the AlN lattice with the concurrent formation of Al vacancies (V$_{\rm Al}$)\cite{7,8}. These are required to absorb the extra valence $e^-$ of O, having an electronic configuration $2{\rm s}^2 {\rm p}^4$, compared to the replaced N with $2{\rm s}^2 {\rm p}^3$. 

In our previous work on polycrystalline sputter deposited films of Al-Si-N, we demonstrated that it is possible to incorporate up to $6\%$ Si into the wurtzite lattice substituting the Al (Si$_{\rm Al}$)\cite{9,10}. Following the arguments of Harris we concluded that the electron donor Si replaces the Al in the AlN structure and one V$_{\rm Al}$ is formed per three Si atoms added into the AlN. An X-ray diffraction (XRD) analysis revealed a shrinking c-axis wurtzite lattice parameter with increasing Si content for 0 to $6\%$ Si. Beyond $6\%$ Si content, the c-axis lattice parameter remained constant\cite{9,10}. The XRD data and a TEM analysis suggested that up to $6\%$ Si can be incorporated into the AlN lattice, while a SiO$_2$ grain boundary phase formed from the extra O surpassing the $6\%$ Si-solubility limit. 
\emph{Ab initio} calculations could reproduce the observed lattice shrinkage under the condition that one V$_{\rm Al}$ was incorporated per three Si$_{\rm Al}$ substitutions, but could not address the mechanisms leading to the experimentally observed Si-solubility limit\cite{11}.

Similarly, in our most recent work, a lattice shrinkage was also observed for increasing O contents up to $8\%$ in sputter-deposited Al-O-N films\cite{Fischer:2019gl}. While our XRD data suggests that up to  $8\%$ O can be incorporated into the AlN wurtzite lattice, Harris et al found that only up to $0.75\%$ O can be incorporated into the wurtzite lattice of single crystalline materials synthesized under thermal equilibrium conditions\cite{5}.

Hence, our previous work performed on sputter-deposited polycrystalline Al-Si-N and Al-O-N films, the work by Harris on Al-O-N single crystals, and the arguments of Slack concerning stochiometry all corroborate that the observed lattice shrinkage arises from vacancies that compensate for excess $e^-$ brought into the initial crystalline structure by aliovalent substitutes. The solubility limit was found to depend strongly on the fabrication method, i.e. is below 1\% for fabrication methods involving high temperatures and up to several percent for sputter deposition methods. This indicates that quenching during fabrication may play an important role. Further, the increasing number of defects will finally destabilize the AlN lattice. XRD and TEM observations indeed revealed that the AlN crystallites shrink with increasing O content and an amorphous grain boundary phase grows in volume. 

The theoretical treatment of such complex materials systems remains challenging: Large 
super-cells have to be used to accommodate a sufficiently large number of foreign atoms and the defects. Consequently, the number of microstates (distributions of the foreign atoms and defects within the crystalline unit cell) grows exponentially which does not permit the optimization of all microstates within Density Functional Theorey (DFT). A human-based selection of tentatively appropriate microstates may thus lead to a misinterpretation of experimental results and may not correctly capture the mutual formation of vacancy clusters and a possible thermodynamical destabilization of the crystallites.

Here we apply \emph{ab initio} Random Structure Searching (RSS)~\cite{pickard2011ab,tan2019structures,rowe2020accurate} combined with advanced high-dimensional structural descriptors and dimensionality reduction to capture different local and global configurational arrangements of the O substitutes and V$_{\rm Al}$.
From this, the expected c-axis parameter, the enthalpy of the super-cell and the vacancy distribution is obtained 
and linked to an intuitive low-dimensional representation of the structure/property relationships. Further, we experimentally address the thermodynamical stability of the solid solution phase of sputtered Al-O-N films using high-temperature annealing experiments and validate the existence and distribution of vacancies predicted by our theoretical approach  by positron annihilation lifetime spectroscopy. While the results obtained here are specific for the Al-O-N system, the theoretical framework proposed remains general and could thus also be applied to evaluate properties of other complex materials systems. 



\section{Experimental and Computational Procedure}

Al-O-N thin film samples were fabricated by reactive direct current magnetron sputtering (R-DCMS) at 200~\degree C, using N$_2$ and O$_2$ as reactive gases.  
O contents between 0.4 and 59.5\% (0\% O corresponding to binary AlN, 60\% O to Al$_2$O$_3$, respectively) were obtained by the adjustment of the O$_2$ flow into the sputter process. 
The O composition was determined by Rutherford Backscattering and refined by Elastic Recoil Detection Analysis (RBS/ERDA).  For the DC sputter deposition of films with high O contents a special gas flow setup was implemented in the sputter chamber to avoid oxygen-poisoning of the Al targets~\cite{13}. The crystallinity of the films, in particular the c-axis lattice dimension, was investigated by symmetric XRD measurements.
As described in detail in our recent work\cite{Fischer:2019gl}, the Al-O-N films show a microstructural evolution with increasing O content that is distinguished by three distinct O concentration regimes, comparable to those found in our earlier work on the Al-Si-N system\cite{9,10}. Al-O-N with $0-8\%$ O, regime (I), forms a crystalline solid solution (Al-O-N$_{\rm ss}$) with a (002) oriented wurtzite fiber texture. Al-O-N with $8-16\%$ O, regime (IIa), consists of a (002) fiber textured nanocomposite in which wurtzite crystallites are gradually encapsulated by an amorphous Al$_2$O$_3$ (a-Al$_2$O$_3$) matrix. Al-O-N with $16-30\%$ O, regime (IIb), forms a nanocomposite without uniaxial texture. At the threshold of $16\%$ O which separates (IIa) from (IIb), the residual film stress changes from tensile (in fiber textured nanocomposites) to compressive (nanocomposites containing crystallites of arbitrary orientations). Al-O-N with $30-60\%$ O, regime (III), consists of an X-ray amorphous solid solution\cite{Fischer:2019gl}.

For the atomistic modeling of the system we concentrate on Al-O-N films containing between 0 and 12 at\% O to analyze regime I for 0 to 8\% O as well as the transition to regime IIa where a (002) fiber texture still exists, but the additional O does not increase the solid solution concentration and forms an amorphous grain boundary phase of increasing thickness. 

As in our earlier work~\cite{11}, we chose a large super-cell for our modeling work consisting of 48 units of the primitive hexagonal cell, and thus containing 192 atoms for the case of pure AlN with no vacancies. This large unit cell permits the incorporation of up to 12 at\% O as well as one V$_{\rm Al}$ per three O to accomodate the extra valence e- arising from the \revision{substitution} of the N by an O. Hence, $3n$ N atoms with $1\leq n \leq 21$ can be replaced by O, and $n$ V$_{\rm Al}$ have to be introduced into the super-cell for the modeling of different O contents. 

The number of possible configurations for such a super cell can be calculated by a simple combinatorial approach

\begin{equation}
    \Omega = 
                    \binom{N_{\rm N}}{n_{\rm O}} \cdot \binom{N_{\rm Al}}{n_{\rm V}} 
                   \,\, ,
\label{eq:eq1}
\end{equation}

where $N_{\rm N}$ is the total number of the nitrogen atoms in the lattice, $N_{\rm Al}$ the number of aluminum atoms, $n_{\rm O}$ is the number of substituional O atoms and $n_{\rm V}$ is the number of V$_{\rm Al}$ (which is defined as $n_{\rm O}/3$). Clearly, even for a small O concentration, a very large number of configurations exist. 
For example, the number of configurations that one can trivially generate by replacing 21 distinct N atoms with O out of the initial 96, and removing 7 Al atoms is 9.33 $10^{30}$. Eq.\,\ref{eq:eq1} further allows the calculation of the configurational entropy of mixing as $\Delta S_{\rm conf}=k_{\rm B}\ln(\Omega)$, a quantity which needs to be considered when analyzing the thermodynmics stability of the system~\cite{25}. DFT modeling all possible microstates is clearly beyond to dates computational possibilities. However, from earlier work~\cite{11} we learned that properties such as the lattice dimensions (and thus also other relevant system parameters) critically depend on the selected microstates. Thus, the  selection of these must be performed without introducing artifacts arising from an arbitrary human-based selection of geometries. 


Here, we rely on RSS combined with a sparsification approach based on a state-of-the-art~\cite{de2016comparing,de+16pccp} global structural metric, followed by the use of dimensionality reduction~\cite{tribello2019using,ceri+13jctc} to infer non-trivial structure/property relationships as discussed in the \emph{Results and Discussion} section. In line with previous RSS works~\cite{pickard2011ab,tan2019structures,rowe2020accurate}, from a large number of distinct random structures we generate a statistically meaningful ensemble of meta-stable states through minimization of an accurate potential energy surface (PES). The RSS begins with the ideal w-AlN lattice, where the indices of N and of Al  atoms are known and can be used to define a set of new configurations by simply replacing groups of N atoms with the same amount of O, removing randomly one Al every 3 O substitutions. We use the function \codeword{combinations} from the Python library \codeword{itertools}~\cite{bernard2016iterators} to generate all the groupings of $\mathcal{X}$ elements from the initial 96 N atoms, with $\mathcal{X}$ being the number O atoms to inject in the lattice, i.e. 3 (1.57\%), 6 (3.16\%), 9 (4.76\%), 12 (6.38\%), 15 (8.02\%), 18 (9.68\%) and 21 (11.35\%). This step results in a large set of possible structures from which we randomly pick 100,000 configurations. For each new structure we randomly remove one Al atom every 3 O atoms.
Finally, for each O concentration, we select ca. 150 structures being the most dissimilar within the set. This can be achieved by having a proper measure of similarity between two structures, for which we use the average Smooth Overlap of Atomic Potential (SOAP) global kernel~\cite{de2016comparing}, as implemented in the \texttt{librascal} package~\cite{librascal}. 
SOAP~\cite{bart+13prb} is a general state-of-art atom-centered, density-based representation of the atomic environment which has been proven very powerful for both properties prediction~\cite{de+16pccp,paru+18ncomm,will+19jcp} and structural classification\cite{helf+19fmb,gasparotto2019identifying}. Extensive details on SOAP is provided elsewhere~\cite{bart+13prb,will+19jcp}. SOAP vectors were computed using a cutoff radius of 5~\AA~, \texttt{n}$_{\rm max}$ = 6 and \texttt{l}$_{\rm max}$ = 6.  The width of the Gaussian functions was set to 0.6~\AA~.  All other parameters were set to their default values. Further details of the analysis of the global similarity analysis between structures will be discussed in the following \emph{Results and Discussion} section.
Having a defined a pairwise similarity measure between global structures, the most different structures can be selected using farthest point sampling (FPS), which has been proven to be well suited for selecting the most widely spread set of landmarks from an initial (larger) set~\cite{ceri+13jctc,tribello2019using}.


For each initial microstate, the equilibrium lattice parameters were computed imposing zero pressure onto the system and optimizing the internal coordinates until atomic forces were lower than 0.01\,eV/\AA~.
The calculations were performed by means of the AiiDAlab platform~\cite{yakutovich2020aiidalab} within DFT as implemented in the CP2K code~\cite{15}. We used the  PBE parameterization for the generalized gradient approximation to the exchange-correlation functional~\cite{14}. Norm-conserving pseudopotentials~\cite{17} were used to describe electronic core-valence interactions.

In addition, the enthalpy $\Delta H_{\rm ass}$ for the association of wurtzite AlN and sapphire Al$_2$O$_3$ to a solid solution of Al-O-N including V$_{\rm Al}$ is calculated from DFT data at a temperature (T) of 0\,K according to the equation w-AlN + $\alpha$-Al$_2$O$_3 \to$Al-O-N$_{\rm ss}$.

\section{Results and Discussion}

\begin{figure}[!bth]
        \centering
        \includegraphics[width=0.55\textwidth]{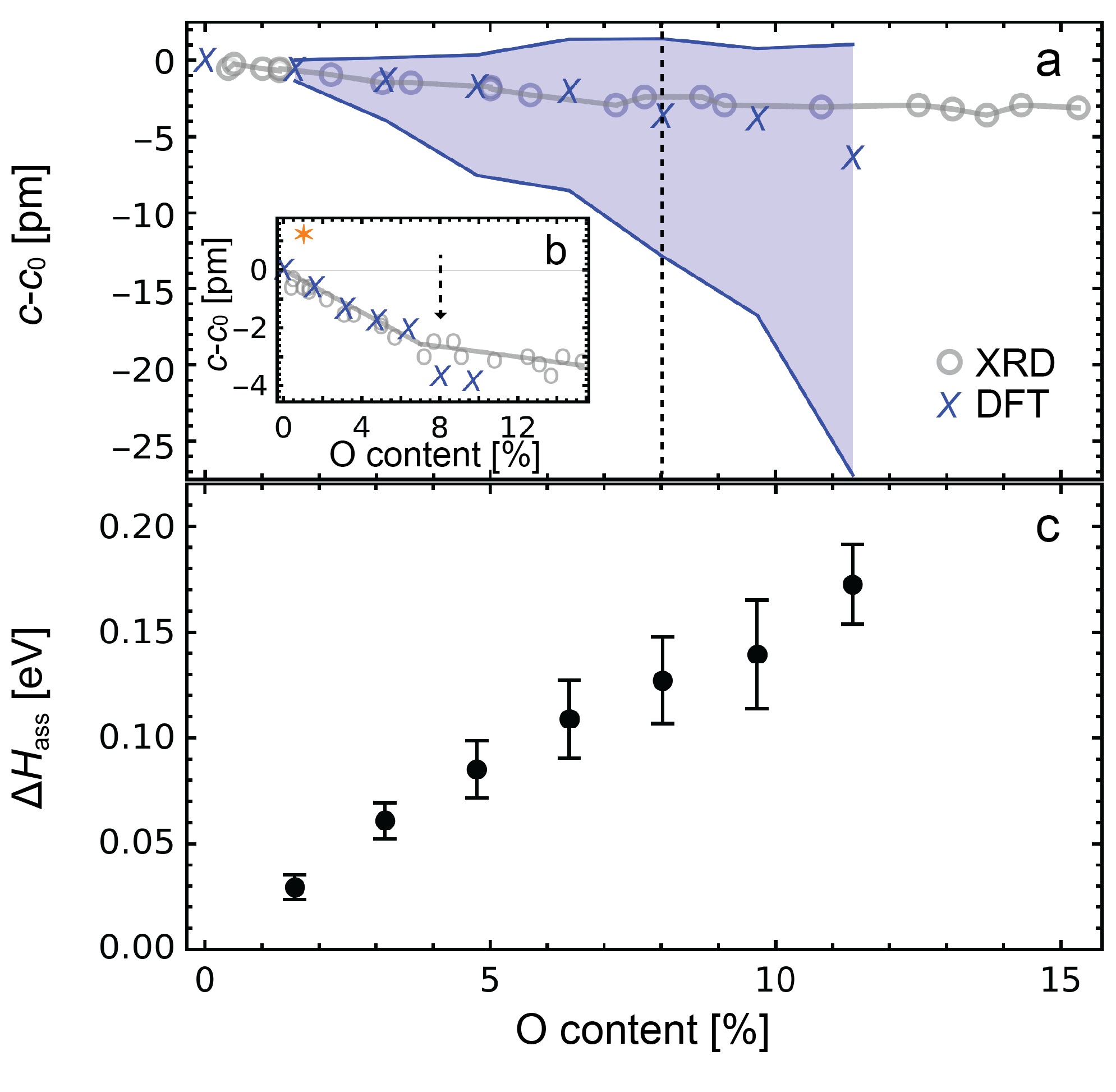}
        \caption {
        Plots of the dependence of the c-axis lattice parameter deviation in Al-O-N from that of AlN on O content. Experimental and theoretical results are plotted by grey open circles and blue crosses, respectively. The orange marker in inset b) represents a DFT-optimized structure without Al vacancies. a) Plot with a wider $c-c_{\rm 0}$ parameter range to display the increasing width of the lattice parameters found for different configurations with increasing O content. The blue crosses show the mean of the $c-c_0$-values obtained from different DFT optimized Al-O-N supercell structures for a specific O concentration. b) Plot of the same data shown in a) with a narrower $c-c_{\rm 0}$ scale to highlight the good agreement of the DFT c-axis lattice values with the experimental results for O concentrations between 0 and 8\%. c) Average $\Delta H_{\rm ass}$ per atom as a function of the O content. Bars represent the standard deviation in $\Delta H_{\rm ass}$ associated with the structural ensemble corresponding to a speciﬁc O concentration.
        }
        \label{fig:figone}
\end{figure}

\begin{figure*}[!bth]
        \centering  
        \includegraphics[width=\textwidth]{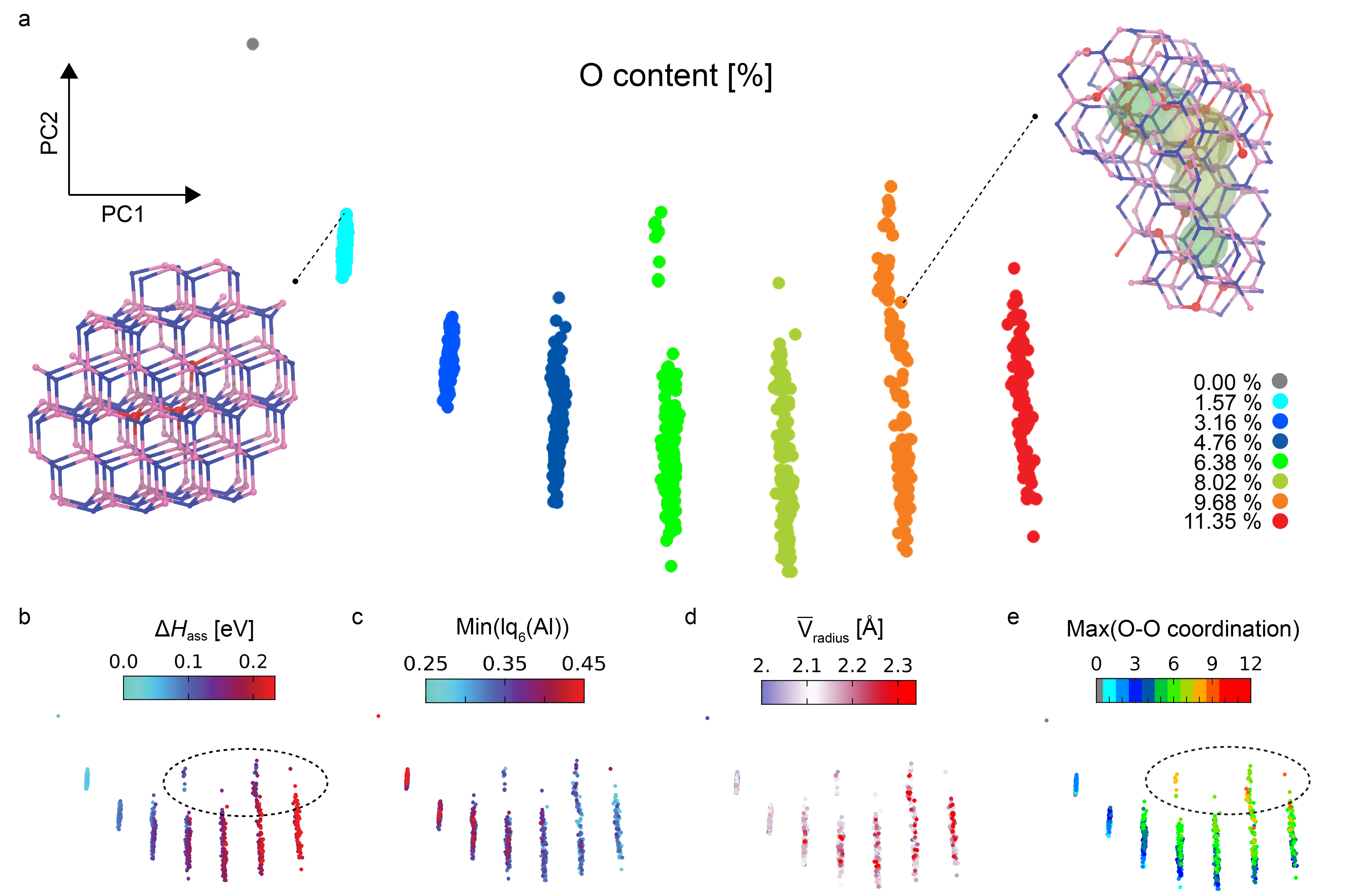}
        \caption {Kernel Principal Component Analysis  map of the global structural similarity of 1088 distinct DFT geometry-optimised Al-O-N structures. a) The first PC ($x$-axis of the map) correlates strongly with the oxygen content within the structure. Vertical stripes correspond, from left to right, to an increased oxygen concentration per structure. Two representative O rich environments are shown for two different O concentrations (left, 1.57\% and right, 9.68\%). The vacancies, highlighted in green, grow bigger for high O concentrations. This can be compared with panel b) where points are coloured according to $\Delta H_{\rm ass}$: structures projected on the right are less stable than those on the left. Other average structural properties can be used as colour code for the map, which helps to link the energetics of the structure with specific structural features. For instance, in panels c), d) and e) points are coloured according to the minimum l$\rm q_{\rm 6}$ order parameter describing the local crystalline order, the average size of voids in the structure and maximum O-O coordination, i.e. the number O neighbors within a cutoff of 3.5~\AA~. Panels d) and e) show that increasing the O concentration, O-rich environment are more likely to be found and this correlates with both the increased void size and the stability ($\Delta H_{\rm ass}$) of the structure. This correlation is captured well by the second PC ($y$-axis), as highlighted by the ellipses. Furthermore the larger the O-O coordination, the shorter the O-O distance as shown in Fig.~\ref{fig:figthree}a. This happens because V$_{\rm Al}$ tend to be coordinated by O atoms, which can get closer to each other when at the void surface.}
        \label{fig:figtwo}
\end{figure*}
Fig.~\ref{fig:figone}a and b replot previous experimental data~\cite{Fischer:2019gl} of the c-axis lattice dependence on the O content in sputter deposited Al-O-N films together with corresponding data obtained from DFT calculations: The c-axis lattice parameter shrinks linearly from 0.498\,nm for pure wurtzite AlN by about 3\,pm for an O content of 8\%. Beyond 8\% O, no further decrease of the lattice parameter was found. We attributed this to the formation of a solid solution of O inside the AlN  grains for up to 8\% O, followed by the formation of an amorphous grain boundary phase for Al-O-N films with higher O contents. Therefore 8\% of O was identified as the solubility limit for the Al-O-N system fabricated by sputter deposition. Making use of the analogies between the Al-Si-N and Al-O-N systems we concluded~\cite{Fischer:2019gl} that in the latter material the observed lattice shrinkage arises from the formation of V$_{\rm Al}$. 

Here we obtain DFT lattice parameter values (depicted as blue crosses in Fig.~\ref{fig:figone}a and b) from 1088 geometry-optimized structures by computing the median c-axis lattice vector at each O concentration. The inset b in Fig.~\ref{fig:figone} shows a zoom around the expectation values to show clearly the different linear trends of the c-axis lattice vector below and above the 8\% solubility limit. The outer box shows the same data with a larger ($c-c_{\rm 0}$)-scale to emphasize the increasing range of lattice parameters obtained for the different configurations with increasing O concentration. The blue shaded area in Fig.~\ref{fig:figone}a represents the range between the minimum and maximum values found within each ensemble. The wide distribution of lattice parameters at different O contents highlights the importance of using a large unit cell and of selecting relevant subsets of configurations as discussed above, also calling for a non trivial interpretation of the results at high oxygen content, way more complex that the simplistic dubbing of “equilibrium solid solution”. The results in panel a and b are complemented by Fig.~\ref{fig:figone}c, which shows that the $\Delta H_{\rm ass}$ per atom increases increasing the amount of O in the structure. Clearly, the solid solution has a higher energy and thus is less favorable than pure AlN and Al$_2$O$_3$.

Our DFT calculations reveal that without the introduction of V$_{\rm Al}$, the lattice expands by 0.25\% when O is introduced, while it shrinks linearly with increasing O content once one V$_{\rm Al}$ every 3 O-atoms is introduced. Note that the latter is required for the compensation of the excess electrons brought in by the O replacing the N. Indeed, when including the V$_{\rm Al}$ DFT results are compatible with the experimental values up to an O content of 8\%. For larger O contents the results obtained from DFT deviate considerably from experimental results. While the lattice parameter obtained from DFT continues to shrink for increasing O contents above 8\%, the experimental data shows no further decrease in the c-axis lattice parameter. This could be attributed to the fact that our DFT calculations take into account the formation of a Al-O-N solid solution only from the incorporation of O into a supercell of single crystalline AlN. The experiment, however, shows a grain refinement and the formation of an amorphous Al$_{\rm 2}$O$_{\rm 3}$ grain boundary phase for O content 8\%. Such a grain boundary phase and the shrinking of the grain size is thus not captured by our DFT models. 
Despite the limits of DFT, insightful information can still be obtained from theoretical data by processing the predicted structures with unsupervised machine learning methods.

Fig.~\ref{fig:figtwo} displays the results obtained by the combination of machine learning methods with advanced descriptors applied to all 1088 structures obtained from DFT for O concentrations ranging from 0 to 11.35\% calculated here. The use of agnostic, high-dimensional descriptors and unsupervised learning allows to shed light on structure-property relationships, 
without restrictions arising a limited number of hypotheses that tend to suffer from a human-bias influenced by the current scientific state-of-the-art knowledge in the field. Each point in the maps a) to g) corresponds to a single Al-O-N structure, with the maps being 2D projections obtained through the application of dimensionality reduction on the pairwise similarity matrix of 1088 geometry-optimized Al-O-N structures. That is to say, the global similarity matrix $M$ is a symmetric 1088x1088 matrix where each component $M_{ij}$ measures the similarity $K(i,j)$ between the pair of structures $i$ and $j$. 
In our case, $K(i,j)$ is a kernel, i.e. a measure of affinity between two different structures. So, if $K(a, b)>K(a, c)$ structures $a$ and $b$ are more similar than structures $a$ and $c$. A kernel must be positive semi-definite and can always be converted into a distance metric. There are many different possibilities to define a kernel function and the suitability of a particular definition typically depends on both the machine learning method being used and on the nature of the data being anlaysed. A simple way to understand the idea of kernel similarity measure is to think about two generic high-dimensional vectors and define the so-called cosine similarity kernel as $k(\vec{x},\vec{y})=\vec{x}\cdot \vec{y}^T/(\norm{\vec{x}}\norm{\vec{y}})$. It is trivial to see that the L2-norm projects the vectors onto the unit sphere, with their dot product being the cosine of the angle between the points denoted by the vectors, i.e. a number smoothly varying between one (aligned vectors) and zero (orthogonal vectors). 

Given the type of data considered here, we define $K(i,j)$ as the average-SOAP kernel between two distinct A-O-N optimized structures $i$ and $j$, as introduced in ref.~\citenum{de2016comparing}. The computation of the pairwise global similarity kernel between the 1088 optimized structures results in a 1088x1088 diagonal matrix that can be eventually interpreted as a structural map by reducing its dimensionality to 2$D$ using kernel principal component analysis (KPCA)~\cite{scholkopf1997kernel}. KPCA is the non-linear form of PCA, which is a well-established technique used to emphasize variation and bring out strong patterns within a dataset. In layman's terms PCA is a basis transformation to diagonalize an estimate of the covariance matrix of a given dataset. The new coordinates in the Eigenvector basis (i.e. the orthogonal projections onto the Eigenvectors) are called principal components (PCs). The selection of the first two PCs generates a 2$D$ map in which the distribution of points qualitatively reproduce the shape of the original high-dimensional manifold. One can then use such a map to colour the points (each point being a structure) according to a specific property, thus providing a simple tool to correlate properties with structures.

Fig.~\ref{fig:figtwo}a is a two-dimensional plot of the DFT configurations where the first and second PC (x- and y-axis) capture the two strongest correlations between different super unit cell configurations obtained by DFT for O contents between 0 and 11.35\%. The color of the points is chosen to represent the O content. Interestingly, there are seven groups of points extending linearly along the y-axis and all points of a group show an identical color. We thus conclude that all configurations relating to a specific O content have a strong similarity, and that the first PCA axis (x-axis) is well correlated with the O content. Note that this finding is not obvious and the widening of the c-axis lattice parameter distribution with increasing O content displayed by the shaded area in Fig.~\ref{fig:figone}a could be interpreted as an increasing dissimilarity of the high O content structures. The unbiased machine learning data analysis approach presented here however clearly documents that all configurations with the same O content have a pronounced similarity.

This finding can be further used to interpret the result shown in Fig.~\ref{fig:figtwo}b, where points are colored according to $\Delta H_{\rm ass}$. The distinct color of the different groups, that according to Fig.~\ref{fig:figtwo}a represent different O contents is compatible 
with the plot of $\Delta H_{\rm ass}$ versus O content displayed in Fig.~\ref{fig:figone}c. Higher O contents clearly lead to higher  $\Delta H_{\rm ass}$-values making these structure energetically less favorable. However, the color distributions of groups attributed to a higher O content tend to widen along the second PCA axis (y-axis). Particularly, the color changes in a rather step-wise fashion along the y-axis. This is particularly apparent for O concentrations between 6.38 and 9.68\%, while the color distribution along the y-axis becomes again more uniform for the highest O content of 11.35\%. We conclude that for O concentrations between 6.38 and 9.68\% two distinct types of configurations must exist, one of them being energetically more favorable, and that a transition from a first type of configuration at low O concentration to a second type existing at the highest O concentration studied here occurs. We note that this spread in the properties of the configurations with higher O content is compatible with the larger spread of the c-axis lattice parameter apparent in Fig.~\ref{fig:figone}a. 

In order to measure the degree of disorder introduced by the O into the wurtzite lattice, we calculate the crystallinity order parameter, i.e. the minimum local $q_{\rm 6}$ ($lq_{\rm 6}$) value for each DFT super cell structure. The $lq_{\rm 6}$ descriptor is a local order parameter designed to distinguish ordered crystal environments from amorphous and liquid environments. Following Ref.~\citenum{li2011homogeneous} we compute, for each atom $i$ within each structure, the quantity

\begin{equation}
     q_{lm}(i) = \frac{1}{N_{\rm b}(i)}\sum\limits_{k=1}^{N_{\rm b}(i)} Y_{ lm}(\theta_{ik},\phi_{ik})\,,
\label{eq:eqlq6}
\end{equation}

where the sum goes over the $N_{\rm b}(i)$ neighbors of the atom $i$, $Y_{ lm}$ are spherical
harmonics, and $\theta_{ ik}$ and $\phi_{ik}$ are the relative orientational angles between the atoms $i$ and $k$. We compute this quantity for all possible values of $\rm m$ and
store them in a vector $\vec{q}_{ l}(i)$ with $2l+1$ components. Finally, we calculate
values $lq_{ l}$ according to

\begin{equation}
    lq_{l}(i) = \frac{1}{N_{\rm b}(i)}\sum\limits_{k=1}^{N_{\rm b}(i)} \frac{\vec{q}_{\rm l}(i) \cdot \vec{q}_{\rm l}(k)}{\abs{\vec{q}_{l}(i)} \abs{\vec{q}_{ l}(k)}}\,.
\label{eq:eqlq6}
\end{equation}

Choosing of $l=6$ has been proven to be a good descriptor capable of detecting the difference between crystalline and amorphous environments, i.e. the lower the value the more amorphous the surrounding structure~\cite{li2011homogeneous,gasparotto2018recognizing}. 

\begin{figure}[!bth]
        \centering
        \includegraphics[width=0.55\textwidth]{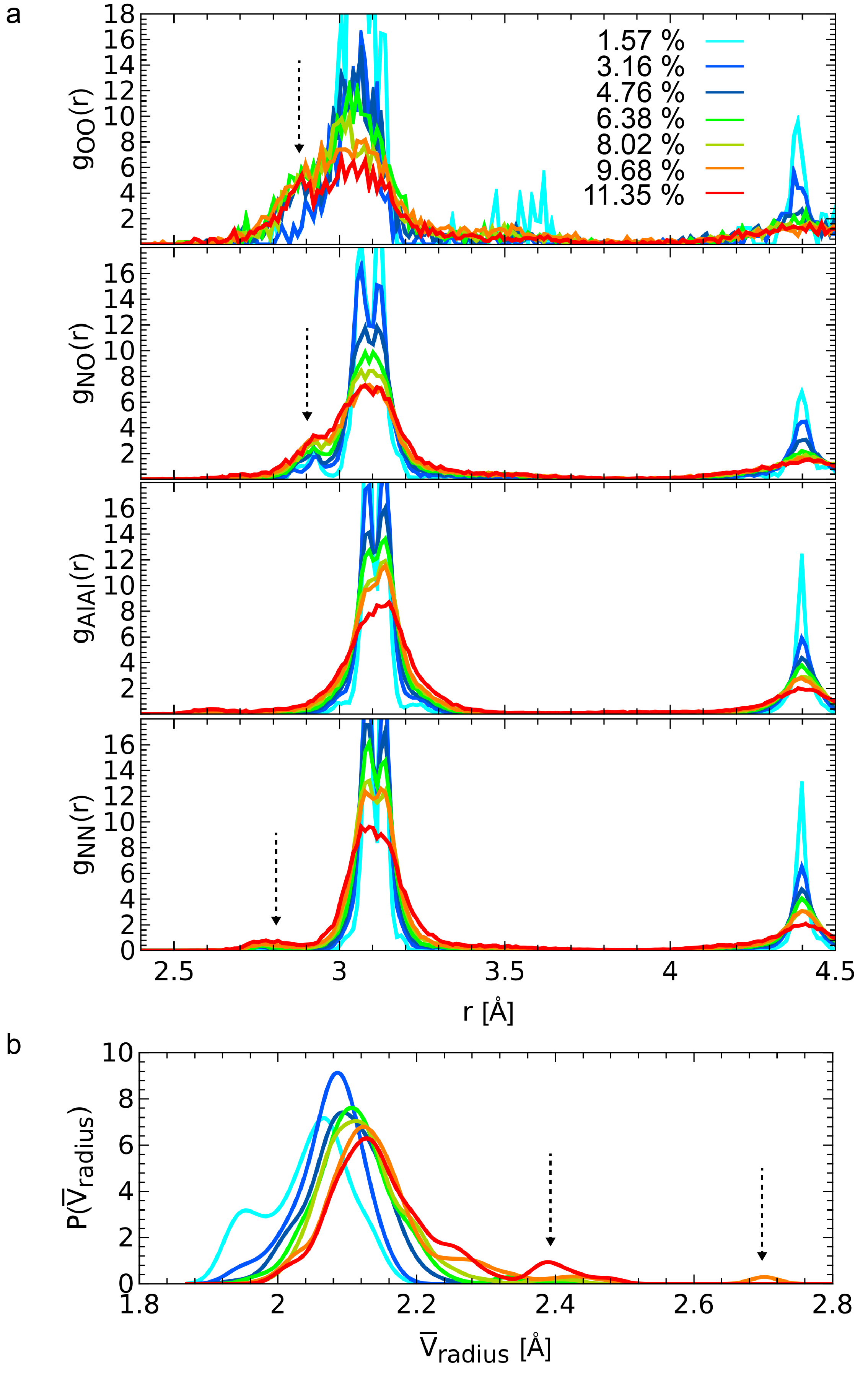}
        \caption {a) Pair-specific radial distribution functions for different O contents. The broadening of the peaks in the pair-correlation functions reveals that the crystalline order decreases with increasing O concentration. Furthermore O-O and N-O show the rise of a small peak (black dashed arrow) at shorter distances for high-O concentrations. This can be attributed to the increased density of O and N atoms near the voids growing in size for increased O concentrations. The vertical arrows indicate the formation of higher densities O-O, N-O and N-N pairs at higher oxygen contents. b) Distributions of void's radii at varying O content. The vertical arrows indicate that voids of larger sizes form at higher O concentrations that further destabilize the Al-O-N lattice.} 
        %
        \label{fig:figthree}
\end{figure}

We then use the minimum $lq_{\rm 6}$ value to color the points arranged in groups in the two-dimensional PC space. From Fig.~\ref{fig:figtwo}c it  appears that the structural disorder increases ($lq_{\rm 6}$ decreases) along the first PC axis, while it varies only little along the second PC axis. This clearly reveals that the structural disorder grows with increasing O content, and that structures with the same O content show similar decreased crystallinity. A correlation with the two groups of $\Delta H_{\rm ass}$ appearing at higher O contents (see Fig.~\ref{fig:figtwo}b) is however not found. Here, the pairwise radial distribution functions (RDFs) displayed in Fig.~\ref{fig:figthree}a provide additional information. The intensities of the peaks at 3.03\,\AA~and 4\,\AA~in the $g_{\rm AlAl}$(r) and $g_{\rm NN}$(r) decays and the peak widths with increasing O content gets broader, confirming the reduction of short range ordering between neighboring atoms compatible with the decreasing $lq_{\rm 6}$ parameter (Fig.~\ref{fig:figtwo}c). The same decay of the peak height (and increased peak width) is also found for the $g_{\rm OO}$(r) and $g_{\rm NO}$(r) distributions. A small additional peak develops for $2.8\leq r\leq 3$\AA~with increasing O content, revealing a growing intensity of closely spaced O atoms and O-N pairs. This development of a second peek in the $g_{\rm OO}(r)$ and $g_{\rm NO}(r)$ distribution functions strongly indicates that two distinct types of structures must exist for higher O content corroborating the two different energy levels apparent by the two distinct colors visible in Fig.~\ref{fig:figtwo}b representing two $\Delta H_{\rm ass}$ energy levels. 

The clustering of O atoms and the increased occurrence of O-N pairs at higher O contents may be accompanied by a clustering of the V$_{\rm Al}$. To quantify this, we plot the void size (computed by using the R.I.N.G.S. package~\cite{ringssw}) and the O-O coordination (the maximum O-O coordination number within a cutoff of 3.5~\AA~ from a central O) in Fig. 2d and e, respectively. The comparison of Fig. 2d with Fig. 2a reveals that the void size generally increases with the O content. 
Moreover, the spread of the colors for a given O content and hence the width of the distribution of the void sizes (Fig. 3b) becomes larger at increased O contents revealing that at higher O contents, different void sizes coexist. The same holds for the Max(O-O)-coordination plotted in Fig. 2e. However, for the high O content structures two distinct Max(O-O)-coordination values appear along the second PC axis (y-axis).  

The machine learning approach used here clearly reveals that a higher O content leads to the formation of vacancy and O clusters apart from a monotonous growth of the association enthalpy $\Delta H_{\rm ass}$. This is in line with other work that has been performed on Al-O-N single crystals which showed that only about 0.5\% of O could be introduced into the AlN wurtzite lattice for AlN single crystals synthesized under conditions remaining close to the thermodynamical equilibrium~\cite{5}.
In stark contrast to the latter work, the much higher O concentration solid solutions obtained in sputtered Al-O-N films~\cite{12}, may possibly arise from an entropic stabilization process: while the growth of Al-O-N single crystals~\cite{5} was performed at high temperatures and a successive slow cool down to room temperature, the thermodynamics occurring during sputtering is substantially different. In sputtering, highly-energetic species bombard the film growing on the substrate. Typical energies of these species (neutrals and ions) are in the range of several eV to a few ten eVs. Consequently the formation of the Al-O-N unit crystallites occurs at an effective temperature that is much higher than the  substrate temperature (here set to 200\degree\,C). In addition, the species deposited onto the substrate leading to the film growth undergo rapid quenching as their kinetic energy dissipated to the thermal bath on a sub-second time scale~\cite{Adamovic:2007gx}. Consequently, a high-temperature state characterized by a high (mixing) entropy could be quenched. We hypothesize that the presence of a sufficiently high-mixing entropy could thus compensate the positive association energy, $\Delta H_{\rm ass}$ such that an entropy-stabilized~\cite{pet+2003,25} solid solution high temperature state could have been quenched to room temperature.

In order to test this hypothesis we designed the following annealing experiments: Al-O-N films with an O content of 10\,\% were sputter-deposited onto Al$_2$O$_3$ substrate. The back side of the substrate was sputter-coated by a 200\,nm film of Ta. XRD of the Al-O-N film (in the front side) revealed a wurtzite (002) peak at 36.2
62$^\circ$ corresponding to a c-axis lattice parameter of 0.496\,nm. The samples were then heated to 1800$^{\circ}$C in vacuum by an \revision{electron} beam arising from a hot filament to the Ta-coated backside of the Al$_2$O$_3$ substrate. The temperature was measured by a pyrometer. At this temperature the mixing entropy stabilizes the solid solution provided the entropy of other phases, e.g. that arising from amorphous grain boundaries, or voids forming in the Al-O-N crystallites remained small. 
After keeping the sample for about 20\,minutes at 1800$^{\circ}$C, the electron beam was shut off. Because of the extreme radiation loss at this high temperature, the sample stops glowing immediately, i.e. cools below a temperature of 1200$^{\circ}$C up to which Al-O-N films were found to be stable, i.e. no atomic rearrangement did occur. We thus expect, that an entropy-stabilized solid solution high-temperature phase would still exist after the rapid cooling process used here. In a further experiment, the system is cooled slowly such that the thermal equilibrium is always kept during cool-down. In this case the entropic contribution to the energy $-T\Delta S$ would become gradually smaller as the sample is cooled such that a Al-O-N$_{\rm ss}$ with high O content would no longer be energetically favorable at lower temperatures and thus would no longer stabilize a solid-solution phase. For this, the samples were heated in an Ar ﬂooded oven and cooled down slowly to room T during a time ramp of 22\,h. 
\begin{figure}[!bth]
        \centering
        \includegraphics{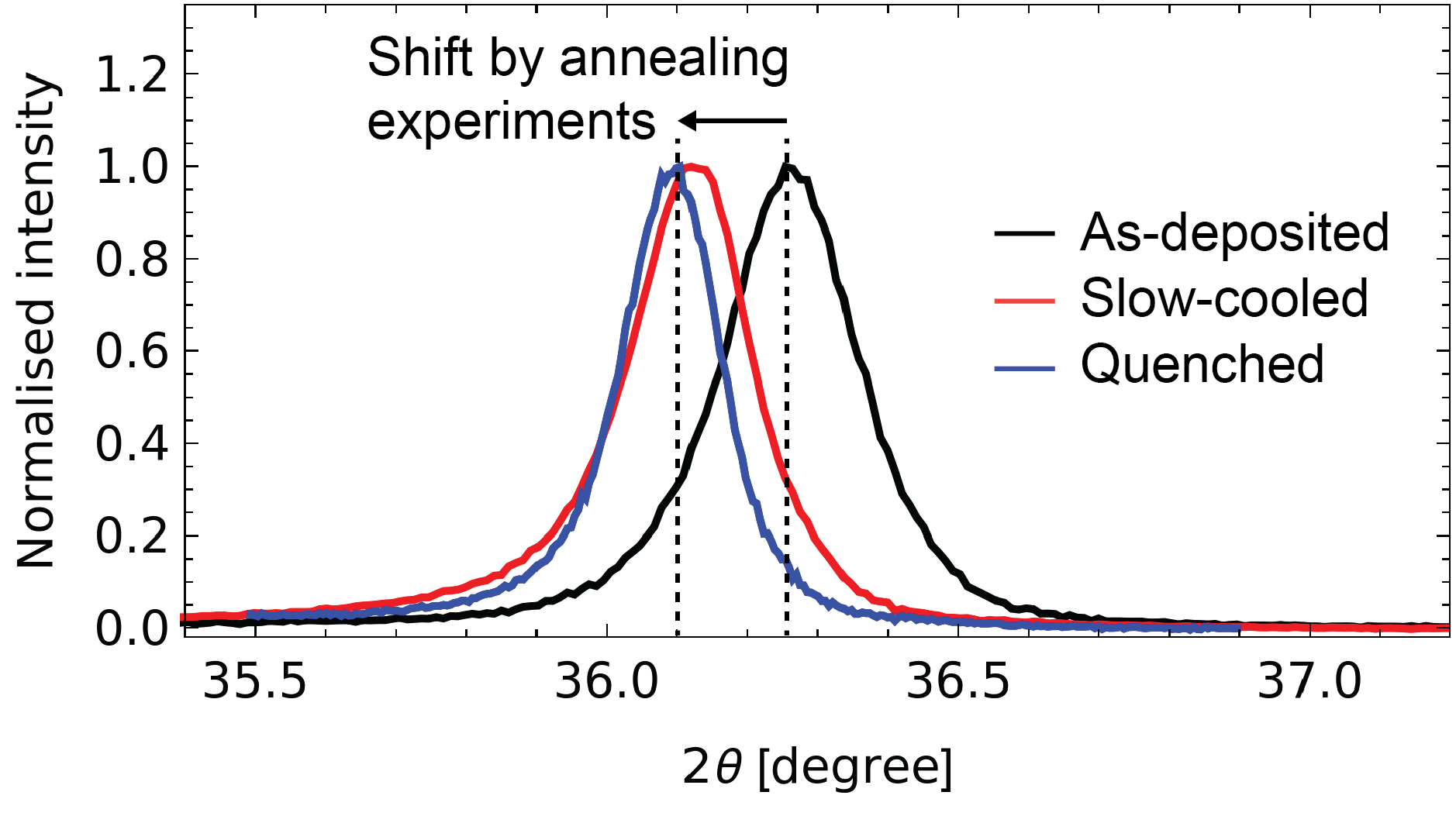}
        \caption {2$\theta$-$\theta$ scans of as-grown and annealed Al-O-N films with oxygen content of around 10 at.\%. After annealing, samples were either slow cooled (red) or quenched (blue). Vertical lines and arrow indicate the shift of the peak positions after annealing.} 
        \label{Fig:entropy}
\end{figure}

Fig.~\ref{Fig:entropy} shows the XRD results obtained from the two experiments. Clearly, both diffraction peaks have shifted from the $36.262^{\circ}$ corresponding to the contracted lattice corresponding to an O content of 10\,\% to about $36.1^{\circ}$ corresponding to the value found for O-free AlN films for both the rapid and slow cooling experiments. An RBS experiment revealed that the initial O content remained constant. We thus conclude that although a sizeable mixing entropy must exist, entropy stabilization of the high-O-content solid solution phase does not occur, and that the O expelled from the AlN grains but still contained in the film has been driven to the grain boundary phase. We speculate that the amorphous grain boundary phase and the formation of voids can be accounted for an entropy higher than the entropy of the phase, such that the thermodynamically stable state consists of AlN grains possibly containing very small amounts of O (below 0.5\%) as found in work on single crystals and an amorphous grain boundary phase~\cite{5}. 

\begin{figure}
\includegraphics{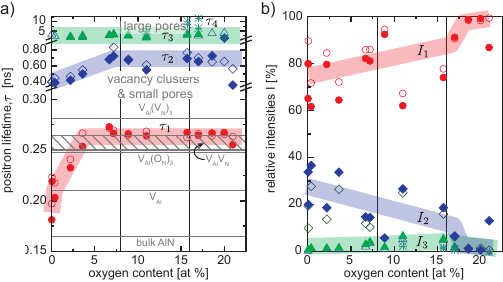}
\caption{\small a) Positron annihilation lifetime components $\tau_1$ (circles),  $\tau_2$ (diamonds), $\tau_3$ (triangles), and $\tau_4$ (stars) as a function of the O content in Al-O-N films. The calculated positron lifetimes for single crystalline AlN, aluminum vacancies, V$_{\rm Al}$ and one V$_{\rm Al}$ arising from 3 O atoms replacing N atoms, a double V$_{\rm Al}$V$_{\rm N}$  (0.253-0.267\,ns)~\cite{Ishibashi:2014fi}, and a V$_{\rm Al}$(V$_{\rm N})_3$ vacancy cluster (0.281\,ns)~\cite{Ishibashi:2014fi} are highlighted by the horizontal grey lines and the hatched \revision{grey} box. The open and full symbols represent data arising from a positron implantation depth of 73 and 180\,nm, respectively. b) plot of the intensities $I_1$ (circles),  $I_2$ (diamonds), $I_3$ (triangles), and $I_4$ (stars) of the lifetime components  $\tau_1$,  $\tau_2$,  $\tau_3$, and  $\tau_3$, respectively, as a function of the O content. The wide transparent lines in panels a) and b) are guides for the eye to highlight the evolution of the different lifetime components and their intensities with O content.}
\label{Fig-PALS}
\end{figure}

Sputtered Al-O-N films containing up to 8\% of O are thus in a metastable state obtained through the the films growth under conditions away from thermal equilibrium. Nevertheless, these thin films are stable up to temperatures of 1200\degree\,C. Even prolonged annealing at such temperatures did not lead to an observable change of the lattice parameter. 
We attribute this to the high melting point of Al$_2$O$_3$ which hinders the kinetics of the O in the metastable Al-O-N$_{\rm ss}$ even at higher temperatures such as 1200\degree\,C. 
Our DFT calculations revealed that this rather stable metastable state is characterized by the formation of O- and V$_{\rm Al}$-clusters at higher O contents accompanied by a lattice instability and amorphization for O contents larger than 8\%, compatible with the experimental observations~\cite{Fischer:2019gl}. In order to experimentally assess the dependence of the nature of the defects on the oxygen content in our sputter-deposited Al-O-N films, variable energy positron annihilation lifetime spectroscopy (VEPALS) measurements were conducted on Al-O-N samples at the Mono-energetic Positron Source (MePS) beamline at HZDR, Germany \cite{Wagner:2017eg,Wagner:2018eh}. Positrons have been implanted into a sample with discrete kinetic energies $E_{\rm p}$ in the range between 0.05 and 10\,keV, which allows for depth profiling from the surface down to several hundred nanometers. A mean positron implantation depth can be approximated by a simple material density dependent formula $z_{\rm mean} = 36/\rho \cdot E_{\rm p}^{1.62}$, where $z_{\rm mean}$ is in nanometers, $\rho = 3.26\,{\rm g}{\rm cm}^{-3}$. For the measurements shown with the open and closed symbols in Fig.~\ref{Fig-PALS} an energy of 3.2 and 5.6~keV providing implantation depths of about 73 and 180\,nm, respectively, were used. Along their tracks inside the Al-O-N film the positrons lose their kinetic energy due to thermalization and after short diffusion annihilate with the electrons emitting two anti-collinear 511\,keV gamma photons that are detected and a positron lifetime spectrum $N(t)$ is recorded. This spectrum is then fitted by a sum of exponentials:

\begin{equation}
N(t) = \sum_{i=1}^n \frac{I_i}{\tau_i}\exp (-\frac{t}{\tau_i}) \,\, ,
\label{eq:lifetime}
\end{equation}

where $\tau_i$ and $I_i$ are the positron lifetime and relative intensity of the i-th component, respectively, and $\sum_{i=1}^nI_i =1$. All the spectra were deconvoluted using the non-linearly least-squared based package PALSfit fitting software~\cite{Olsen:2007gw} into three or four discrete lifetime components that are characteristic for different defect types and sizes. 

In Fig.~\ref{Fig-PALS}a) and b) the different positron annihilation lifetime components and the corresponding intensities are plotted as a function of oxygen content, respectively. The good agreement of the data for implantation depths of 73 and 180\,nm (displayed with open and full symbols, respectively) indicates that the defect concentration and defect types do not depend on the depth inside the Al-O-N film compatible with the morphology of the films mapped by TEM in our earlier work\cite{Fischer:2019gl}. For comparison, the positron annihilation lifetimes calculated for bulk AlN (0.158\,ns)~\cite{Tuomisto:2008bx}, Al-vacancies, V$_{\rm Al}$ (0.208\,ns)~\cite{Tuomisto:2008bx,Maki:2011jk}, for a single Al-vacancy arising from three oxygen atoms replacing three nitrogen atoms, V$_{\rm Al}$(O$_{\rm N}$)$_3$ (0.241\,ns)~\cite{Ishibashi:2014fi}, a double V$_{\rm Al}$V$_{\rm N}$  (0.253-0.267\,ns)~\cite{Ishibashi:2014fi}, and a V$_{\rm Al}$(V$_{\rm N})_3$ vacancy cluster (0.281\,ns)~\cite{Ishibashi:2014fi} are displayed by means of horizontal grey lines and the hatched grey box in Fig.~\ref{Fig-PALS}a. The lifetime component $\tau_1$ (open and closed circles) 
is compatible with the existence of V$_{\rm Al}$ and V$_{\rm Al}$ surrounded by O$_{\rm N}$ clusters. The linear rise of $\tau_1$ from about 0.188 to 0.262\,ns for an oxygen concentration between 0 and 3.7\,\% is an indication for the formation of larger defects, e.g. divacancies. The second lifetime component $\tau_2$ (open and close diamonds) that rises linearly from 0.4\,ns to about 0.7\,ns can be  attributed to formation of vacancy clusters and pores with a diameter increasing from about $0.45-0.48\,$nm\cite{Ghasemifard:2020bg} for an O content increasing to 8\,\%. Such an increase of the pore diameter for higher oxygen contents is compatible with 
our unsupervised machine learning analysis of the DFT data, which revealed an increase of the vacancy radius from about 2.1\,{\AA} to 2.4\,{\AA} at O concentrations increased from a few percent to about 8\,\% (Fig.~\ref{fig:figtwo}d) and the peaks at larger radii appearing in the plots displayed in Fig.~\ref{fig:figtwo}b.
The PALS results for lower O contents must however have a different origin not captured by our DFT results that consider only an Al-O-N supercell, but not the true textured multigrain film structure: We attribute the existence of the second (and third) lifetime components at low O contents to the open grain boundaries that also lead to a tensile film strain\cite{Fischer:2019gl}. 

The lifetime intensities shown in Fig.~\ref{Fig-PALS}b show a significant correlated scattering of the $I_1$ and $I_2$ lifetime intensities. We attribute this to the local variation of defect micro-structure (most probably grain boundaries) for certain O concentrations, with however the major defect types remaining unchanged. In-spite of the large correlated scattering an increase of $I_1$ is apparent, while $I_2$ decreases (as highlighted by the wide red and blue trend lines, respectively). The reduction of the intensity $I_2$ of the second lifetime component $\tau_2$ with increasing O concentration together with the existence of $\tau_2$ at low O concentrations highlights the two distinct origins that give rise to the second lifetime component: At very low O concentrations, no large voids exist inside the Al-O-N grains, but the morphology of the films is governed by open grain boundaries that consequently give rise to longer positron annihilation lifetimes. At increased O contents our previous work\cite{Fischer:2019gl}, revealed that a grain refinement occurs and an amorphous grain boundary phase forms that is gradually closing up the open grain boundaries and consequently reduces the tensile film strain. We thus attribute the observed reduction in $I_2$ to a corresponding reduction of the open grain boundary volume for increasing O concentrations. Interestingly for an O concentration of about 16\,\% O, seven out of eight $I_2$ data points indicate a rapid drop of $I_2$ towards zero while the corresponding $I_1$ data points raise to almost 100\% (see blue and red trend lines in Fig.~\ref{Fig-PALS}b). This observation is compatible with the morphological transition occurring at an O content of 16\% established in our previous work\cite{Fischer:2019gl} which showed a transition from tensile to compressive strain compatible with the absence of open grain boundaries and the formation of a compact nanocrystalline phase. No clear interpretation is possible for the third and forth lifetime components, even though they could represents a low concentration of local voids in amorphous Al$_2$O$_3$ phase, which attract positrons.

\section{Conclusions}
Sputter-deposited Al-O-N films (and similar systems such as for example A-Si-N) fabricated by sputter-depostion are a surprisingly complex class of materials~\cite{12} with potential applications such as protective transparent coatings. In earlier work we found that the Al-O-N~\cite{12} (and also Al-Si-N films~\cite{9,10}) undergoes a complex structural evolution with increasing O content. At lower O (and Si contents) XRD indicated the formation of a solid solution by integration of the O (and Si) into the AlN wurtzite lattice. In earlier theoretical work, the Si atoms in the Al-Si-N films were reasonably distributed in the AlN wurtzite host lattice for DFT calculations. The results confirmed the lattice contraction measured by XRD. However, here we show that the human-choice of selecting solely super cells with nicely distributed V$_{\rm Al}$ and O atoms (or Si atoms in our earlier work) is questionable, particularly when considering films fabricated by sputtering which is a highly off-equilibrium process.
RSS is free from human predjustice and allows for fast generation of a huge amount of configurations than can be reduced to small set of initial candidates using FPS on a smart metric to compare structural pairs. Accurate DFT can then be applied to obtain stable structures from the initial set of ideal configurations. The results from DFT then need to be further interpreted to shed light on the link between structure and properties, which is again challenging and is performed here again without any human bias by exploiting the power on unsupervised learning. Explaining experiments surprisingly complex, since the interpretation of the XRD data of the Al-O-N (and similar systems, e.g. Al-Si-N) shows a linear contraction of the lattice with an O content increased from 0 to 8\%, while it remains constant for higher O contents, making the interpretation challenging. Based on the work presented here, we suggest to refrain from naming an Al-O-N system with high O contents of a few percent a \emph{solid solution} of O inside the AlN wurtzite lattice (and the same statement would apply to many other similar materials systems). Indeed, solid solution implies that O and V$_{\rm Al}$ are well distributed inside an otherwise perfect AlN wurtzite lattice and that the material is in a thermodynamically stable state. This is however not the case. The theoretical results presented here revealed an increasing density of larger vacancy clusters (i.e. voids, confirmed by PALS), with the formation of O and O-N pairs which strongly distorts the AlN lattice in the proximity of these large defects. The large spread in the observed vertical lattice constant found via DFT calculations across the large set of investigated configurations is also an indicator of a complex scenario that had been oversimplified in previous interpretations.

Hence, sputtered Al-O-N films with high O contents can be better understood as a mixture of small metastable Al-O-N grains including larger scale defects such as V$_{\rm Al}$ clusters voids, surrounded by an O and N-rich phase and leading to strong local lattice distortions. Possibly, these defect structures can accommodate further O introduced into the system such that the lattice parameter of the metastable Al-O-N grains that shrink in size with increased O contents would remain constant as detected by XRD. We anticipate that our findings lay the foundations for future investigations devoted to studying the structural stability of defective sputter-deposited materials. Indeed, recent achievements on the development of robust machine learning-based atomistic simulation frameworks~\cite{schutt2018schnet,wang2018deepmd,bartok2010gaussian,behler2007generalized} allow now for training next-generation machine learning potentials capable of extending our theoretical model up to millions of atoms~\cite{lu202086} while retaining \emph{ab initio} accuracy. This would allow probing the energetic of much larger supercells, with the possibility of testing the hypothesis of grain boundaries formation.

We conclude that sputter deposited Al-O-N with O concentrations up to a few percent are in metastable state having a high O concentration in the AlN wurtzite lattice. The incorporation of the O is accompanied by the formation of V$_{\rm Al}$ to accommodate the extra valence $e^-$ of O compared to N. At O concentrations of a few percent, the formation of larger and V$_{\rm Al}$-O clusters is observed which lead to an increasing destabilization of the AlN wurtzite lattice. Although sputter-deposited Al-O-N with high O contents in the AlN wurtzite lattice are surprisingly stable up to temperatures of about 1200$
^\circ$C, 
their thermodynamically stable microstructure is a nanocomposite, in which (almost) O-free wurtzite AlN grains are surrounded by amorphous Al$_2$O$_3$ grain boundary phase.

\begin{acknowledgement}
The fabrication of the Al-O-N thin films, and their XRD
characterization has been performed within the Swiss National Science Foundation 
project No. 200021\_150095 that is acknowledged here. 
The PALS measurements were carried out at ELBE at the Helmholtz-Zentrum Dresden-Rossendorf e. V., a member of the Helmholtz Association. We would like to thank Ahmed G. Attallah and Eric Hirschmann for assistance. This work was partially supported by the Impulse-und Networking fund of the Helmholtz Association (FKZ VH-VI-442 Memriox), and the Helmholtz Energy Materials Characterization Platform (03ET7015).
\end{acknowledgement}




\bibliography{achemso}

\end{document}